\documentclass[a4paper,12pt]{elsart}
\usepackage{geometry}
\usepackage{amssymb}
\journal{Physics Letters A}
\begin{document}
\begin{frontmatter}
\title{Statistical theory of self-similar time series \\ as a nonextensive
thermodynamic system}
\author{Alexander I. Olemskoi\thanksref{AO}}
\address{Max-Planck-Institute f\"ur Physic komplexer System,
N\"othnitzer Strasse 38, D-01187 Dresden, Germany}
\thanks[AO]{Corresponding address: Sumy State University,
Rimskii-Korsakov St. 2, 40007 Sumy, Ukraine. E-mail: olemskoi\char'100ssu.sumy.ua}

\begin{abstract}
Within Tsallis' nonextensive statistics, a model is elaborated
to address self-similar time series as a thermodynamic system.
Thermodynamic-type characteristics relevant to temperature, pressure, entropy, 
internal and free energies are introduced and tested. 
Stability conditions of time series analysis are discussed in details 
on the basis of Van der Waals model.
\end{abstract}

\end{frontmatter}

\section{Introduction}

Time series analysis allows to elaborate and verify macroscopic
models of complex systems behaviour on the basis of data analysis.
The latter is known to be focused on numerical calculations of
correlation sum for delay vectors that allow to find relevant
magnitudes of fractal dimension and entropy \cite{1}. Being
traditionally a branch of the theory of statistics, time series
analysis is based on the class of models of harmonic oscillator,
which are relevant to the simplest case of the Gaussian random
process \cite{A}. But it is well known that real time series is
rather the L\'evy stable processes related to self-similar system,
than the Gaussian ones being very special case \cite{B}.
Therefore, it is important to study self-similar time series. In
this Letter we present a method based on Tsallis' statistics
\cite{2}, which is of relevance for self-similar system analysis
\cite{3} -- \cite{6}.

The paper is organized as follows. Section 2 is devoted to
elaboration of a model, which permits to address a self-similar
time series as a nonextensive thermodynamic system. 
Section 3 is based on calculations of both entropy and internal energy of the time series. 
As a result, thermodynamic-type characteristics of the time series such as
temperature and entropy, volume and pressure, internal and free energies are introduced. 
Their testing for the model of ideal gas is shown to be
basis for statistics of self-similar time series. 
Section 4 is devoted to discussion of the physical meaning of the results obtained. 
Stability conditions of time series mimicked as a non-ideal gas are discussed in details when 
external field and particle interaction are switched on.
Several equalities needed in quoting are placed in Appendix.

\section{Presenting time series as nonextensive thermodynamic system}

Let us consider $D$-dimensional time series ${\bf x}(t_n)$ related
to the set $\{{\bf x}_n\}$ of consequent values ${\bf
x}_n\equiv{\bf x}(t_n)$ of principle variable ${\bf x}(t)$ taken
at discrete time instants $t_n\equiv n\tau$ that we obtain as
result of dividing a whole time series length $\mathcal{T}\equiv
N\tau$ by $N$ equal intervals $\tau$. It is obviously to be relevant
to the time series ${\bf x}(t_n)$, the set $\{{\bf x}_n\}$ should
be supplemented by conjugated set $\{{\bf v}_n\}$ of velocities, which show
rates of ${\bf x}_n$-variation with the time jumping. In the
simplest case of Markovian consequence, one has ${\bf v}_n\equiv
({\bf x}_{n}-{\bf x}_{n-1})/\tau$. For more complicated series
with $m$-step memory, the velocity magnitude is defined as
follows:
\begin{eqnarray}
v_n\equiv\left[
{1\over m}\sum\limits_{i=1}^{m}{\vec \delta}_i^2(m, n)\right]^{1\over 2},\quad
{\vec \delta}_i(m, n)\equiv {{\bf x}_{(n-m)+i}-{\bf x}_{(n-m)+(i-1)}\over\tau}.
\label{A}
\end{eqnarray}

The paradigm of our approach is to address the time series as a physical
system defined by
an effective Hamiltonian $\mathcal{H}=\mathcal{H}\{{\bf x}_n, {\bf v}_n\}$,
on whose basis statistical characteristics of this series could be found.
If one proposes that series terms ${\bf x}_n$ related to different $n$
are not connected, the effective Hamiltonian is additive:
\begin{eqnarray}
\mathcal{H}=\sum\limits_{n=1}^{N}\varepsilon_n,\quad \varepsilon_n\equiv({\bf x}_n, {\bf v}_n).
\label{B}
\end{eqnarray}
Physically, this means that the series under consideration is relevant
to an ideal gas
comprising of $N$ identical particles with energies $\varepsilon_n$.
Further, we suppose different terms of time series to be statistically
identical, so that effective particle energy
does not depend on coordinate ${\bf x}_n$: $\varepsilon({\bf x}_n, {\bf v}_n)
\Rightarrow\varepsilon({\bf v}_n)$.
Moreover, since this energy does not vary with inversion of the coordinate
jumps
${\bf x}_{n}-{\bf x}_{n-1}$, the function $\varepsilon({\bf v}_n)$ should be even.
We use the simplest square form
\begin{eqnarray}
\varepsilon_n={1\over 2}{\bf v}_n^2,
\label{C}
\end{eqnarray}
which is reduced to the usual kinetic energy for a particle with
mass $1$.
With switching on an external force ${\bf F}={\bf const}$, particle energy (\ref{C}) 
becomes as follows: 
\begin{eqnarray}
\varepsilon_n={1\over 2}{\bf v}_n^2-{\bf F}{\bf x}_n.
\label{CC}
\end{eqnarray}
Finally, when time series has a microscopic memory, 
dimension of the delay vectors ${\vec \delta}_i(m, n)$ in the definition (\ref{A}) needs taking $m>1$. 
Moreover, if time series terms ${\bf x}_m$, ${\bf x}_n$ with $m\ne n$ are clustered, the Hamiltonian 
becomes relevant to a non-ideal gas with interaction $w_{mn},~m\ne n$:
\begin{eqnarray}
\mathcal{H}={1\over 2}\sum\limits_{n=1}^{N}{\bf v}_n^2+{1\over 2}\sum\limits_{m\ne n}w_{mn}.
\label{CCC}
\end{eqnarray}

Let a physical quantity $\mathcal {O}$ be a random variable with average
\begin{eqnarray}
\left< \mathcal{O}\right>_q=\sum\limits_{\{{\bf x}_n, {\bf v}_n\}}{\mathcal{O}
\{{\bf x}_n, {\bf v}_n\}}\mathcal{ P}_q\{{\bf x}_n, {\bf v}_n\}
\label{Ccc}
\end{eqnarray}
defined by relevant functional $\mathcal{ O}\{{\bf x}_n, {\bf v}_n\}$.
Hereafter, the principle role is devoted to the sum over states, which reads:
\begin{eqnarray}
\sum\limits_{\{{\bf x}_n, {\bf v}_n\}}\Rightarrow{1\over N!}\int\int\prod\limits_{n=1}^{N}{{\rm
d}{\bf x}_n{\rm d}{\bf v}_n\over\Delta}=
{1\over N!}\left({X^2\over\tau\Delta}\right)^{DN}
\prod\limits_{n=1}^{N}\int{\rm d}{\bf y}_n\int{\rm d}{\bf u}_n
\Rightarrow \mathcal{ N}^{-1}\prod\limits_{n=1}^{N}\int{\rm d}{\bf y}_n\int{\rm d}{\bf u}_n;\quad
\label{D1}\\
\mathcal{ N}\equiv N!\left({X^2\over\tau\Delta}\right)^{-DN}\simeq
\left[{{\rm e}X^{2D}\over N(\tau\Delta)^D}\right]^{-N};\quad
{\bf y}_n\equiv{{\bf x}_n\over X},~{\bf u}_n\equiv{\tau{\bf v}_n\over X},~~
\int{\rm d}{\bf y}_n=\int{\rm d}{\bf u}_n=1.\quad
\label{D2}
\end{eqnarray}
Here, $\Delta$ is effective Planck constant that determines a
minimal volume of phase space per a particle related to a term of
the time series, the factorial takes into account statistical
identity of these terms, the factor $(X^2/\tau)^{DN}$ is
caused by change of initial variables ${\bf x}_n, {\bf v}_n$ into
${\bf y}_n, {\bf u}_n$ rescaled with respect to macroscopic length $X$.

At a choice of the probability distribution $\mathcal{ P}_q\{{\bf x}_n,
{\bf v}_n\}\Rightarrow\mathcal{ P}_q\{{\bf y}_n, {\bf u}_n\}$,
we are stated on a self-similarity of the
time series meaning a power-law form of the functional
$\mathcal{ P}_q\{{\bf y}_n, {\bf u}_n\}$ instead of usual exponential one type of Gaussian.
We adopt known Tsallis' escort distribution \cite{2}
\begin{eqnarray}
\mathcal{ P}_q\{{\bf y}_n, {\bf u}_n\}=\left\{
\begin{array}{ll}
0  &\textrm{at}\quad
(1-q){\mathcal{ H}\{{\bf y}_n, {\bf u}_n\}-E\over\left<1\right>_q T_0}>1,\\
{1\over Z}\left[1-(1-q){\mathcal{ H}\{{\bf y}_n, {\bf u}_n\}-E\over
\left<1\right>_q T_0}\right]^{q\over 1-q}
& \textrm{otherwise}.
\end{array} \right.
\label{D}
\end{eqnarray}
Here partition function is defined by condition
\begin{eqnarray}
Z\equiv \mathcal{N}^{-1}\prod\limits_{n=1}^{N}\int\left[1-(1-q){\mathcal{ H}\{{\bf y}_n, {\bf u}_n\}-E\over
\left<1\right>_q T_0}\right]^{q\over 1-q}{\rm d}{\bf y}_n{\rm d}{\bf u}_n,
\label{E}
\end{eqnarray}
where $0<q<1$ is a parameter of nonextensivity, $T_0$ is energy
scale, internal energy $E$ and normalization parameter
$\left<1\right>_q$ are determined by equalities
\begin{eqnarray}
&&E\equiv \mathcal{ N}^{-1}\prod\limits_{n=1}^{N}\int\mathcal{H}
\{{\bf y}_n, {\bf u}_n\}\mathcal{ P}_q\{{\bf y}_n, {\bf u}_n\}{\rm d}{\bf y}_n{\rm d}{\bf u}_n,\nonumber\\
&&\left<1\right>_q\equiv\mathcal{ N}^{q}\left[\prod\limits_{n=1}^{N}\int\limits_0^1
\left(\mathcal{ P}_q\{{\bf y}_n, {\bf u}_n\}\right)^{1\over q}{\rm d}{\bf y}_n{\rm d}{\bf u}_n\right]^{-q}.\quad
\label{F}
\end{eqnarray}

To check the statistical scheme
proposed let us address firstly trivial case of time series
${\bf x}_n={\rm{\bf const}}$.
Here, the particle energy $\varepsilon$ is a constant as well,
so that the Hamiltonian is $\mathcal{ H}=N\varepsilon$.
The partition function $Z=\mathcal{ N}^{-1}$ and the
normalization parameter
$\left<1\right>_q=\mathcal{ N}^{-(1-q)}$
are given by inverted normalization factor (\ref{D2}), whereas
the internal energy $E=N\varepsilon$ is reduced to the Hamiltonian.
Then, the entropy $H=-\ln\mathcal{ N}$ obtained according to definition
(\ref{H1}) given in Appendix
is reduced to zero if only the normalization factor takes value $\mathcal{
N}=1$.
As a result, we find effective Planck constant:
\begin{eqnarray}
\Delta=\left({{\rm e}\over N}\right)^{1\over D}{X^2\over\tau}.
\label{I}
\end{eqnarray}
If $D$-dimensional domain of the coordinate ${\bf x}_n$-variation supposes to be governed
by L\'evy-type law
\begin{eqnarray}
X^D=x^D N^{1\over z}
\label{J}
\end{eqnarray}
with microscopic constant $x$ and dynamic exponent $z$, then one
obtains the following scaling relation for the phase space volume
per a term of the time series:
\begin{eqnarray}
\Delta^D={\rm e}\left({x^2\over\tau}\right)^{D}N^{{2\over z}-1}.
\label{K}
\end{eqnarray}
In the case of Gaussian scattering, when $z=2$, this volume does not
depend on number $N$ of the time series terms,
i. e., one obtains a constant.

\section{Non-extensive thermodynamics of time series as an ideal gas}

Stating on the basis of known results for thermodynamic quantities
of nonextensive ideal gas
(see Appendix) we are in position to examine pseudo-thermodynamic
properties of time series.
The quoted equalities (\ref{S5}) -- (\ref{G}) derive to the starting
expression for the normalization quantity
\begin{eqnarray}
\left<1\right>_q=\left\{ {X^{DN}\gamma(q)\over N!}
\left[{\theta(1+ a)^{{q\over a}+1}\over 1-q}\right]^{DN\over 2}\right\}^{1-q\over
1- a}.
\label{L}
\end{eqnarray}
Here the notations are introduced (cf. Eqs. (\ref{S7}))
\begin{eqnarray}
\theta\equiv{2\pi T_0\over\Delta^2},\quad
\gamma(q)\equiv{\Gamma\left({1\over 1-q}\right)\over\Gamma\left({1\over
1-q}+{DN\over 2}\right)},\quad
a\equiv (1-q){DN\over 2},
\label{M}
\end{eqnarray}
where we use $\Gamma$-function. In the limits $N\gg 1$, $(1-q){D\over 2}\ll 1$,
when
\begin{eqnarray}
\gamma(q)\simeq\left[{\rm e}(1-q)\right]^{DN\over 2}(1+ a)^{-{1+a\over 1-q}},
\label{N}
\end{eqnarray}
one obtains for the entropy (\ref{H1}):
\begin{eqnarray}
H\simeq{Na\over 2(1- a)}\ln\left[{\rm e}^{2+D}\theta^D\left({X^D\over N}\right)^{2}\right].
\label{O}
\end{eqnarray}
With accounting scaling relation (\ref{J}), this expression takes the usual form
\begin{eqnarray}
H=N{z-1\over z}\ln\left({G\over N}\right),\quad
G\equiv(2\pi{\rm e}T_0)^{Dz\over 2}\left(x\over\tau\right)^{-Dz},
\label{P}
\end{eqnarray}
if the dynamic exponent is determined as
\begin{eqnarray}
z={1\over 1- a}.
\label{Q}
\end{eqnarray}
Respectively, the internal energy (\ref{S6}) and the normalization
parameter (\ref{L}) read:
\begin{eqnarray}
E={DN\over 2}\left({G\over N}\right)^{{2\over D}\left(1-{1\over z}\right)}T_0,\quad
\left<1\right>_q=\left({G\over N}\right)^{{2\over D}\left(1-{1\over z}\right)}.
\label{S}
\end{eqnarray}

Following \cite{9}, let us introduce the temperature
\begin{eqnarray}
T\equiv\left<1\right>_q T_0=\left({G\over N}\right)^{{2\over D}\left(1-{1\over z}\right)}T_0,\quad 
a\equiv{z-1\over z}, 
\label{Ss}
\end{eqnarray}
where the second equality takes into account the last relation (\ref{S}). 
This definition guarantees the equipartition law 
\begin{eqnarray}
E=CT,\quad C\equiv cN,\quad c\equiv{D\over 2},
\label{1}
\end{eqnarray}
where the quantity
\begin{eqnarray}
C={\partial E\over\partial T}
\label{1a}
\end{eqnarray}
is the specific heat. 
It is easily to convince that equations (\ref{P}) -- (\ref{Ss}) arrive at standard thermodynamic relation 
\begin{eqnarray}
{\partial H\over\partial E}\equiv{1\over T}.
\label{33}
\end{eqnarray}

Above used treatment is relevant to the method \cite{7a} addressed to 
a fixed value of the internal energy $E$. 
In alternative case when the principle state parameter is the temperature $T$,
we should pass to the conjugate formalism \cite{7}. Here, standard definition
\begin{eqnarray}
F\equiv E-TH
\label{3aa}
\end{eqnarray}
of the free energy arrives at the dependence
\begin{eqnarray}
F=-CT\ln\left({~T\over{\rm e}T_0}\right).
\label{Zz}
\end{eqnarray}
Then, the thermodynamic identity 
\begin{eqnarray}
{\partial F\over\partial T}\equiv - H
\label{3}
\end{eqnarray}
yields the relation
\begin{eqnarray}
H=C\ln\left({T\over T_0}\right)
\label{Zz1}
\end{eqnarray}
that plays a role of a heat equation of state.
It arrives at the usual definition of the specific heat (cf. Eq. (\ref{1a}))
\begin{eqnarray}
C=T{\partial H\over\partial T}.
\label{1aa}
\end{eqnarray}

Let us introduce now specific entropy per unit time
\begin{eqnarray}
h\equiv(D\tau)^{-1}{{\rm d}H\over{\rm d}N}=
\tau^{-1}H_1-r
\label{S1}
\end{eqnarray}
to be determined by a minimal entropy
\begin{eqnarray}
H_1=z\ln\sqrt{2\pi{\rm e}T_0}-z\ln\left(x\over\tau\right)
\label{S2}
\end{eqnarray}
and a redundancy
\begin{eqnarray}
r=(D\tau)^{-1}\ln({\rm e}N).
\label{S3}
\end{eqnarray}
Dependencies on the scale $x$
\begin{eqnarray}
h(x)={\rm const}-{z\over\tau}\ln\left({x\over\tau}\right),\quad
r(x)={\rm const}
\label{S4}
\end{eqnarray}
notice that the system behaves in a stochastic manner \cite{10}.

Let an effective pressure be defined as 
\begin{eqnarray}
p\equiv-\tau{\partial h\over\partial x}
\label{3a}
\end{eqnarray}
to measure specific entropy variation with respect to the time series scale.
Then, we arrive at a mechanic equation of state
\begin{eqnarray}
px=z
\label{3b}
\end{eqnarray}
being additional to the relation (\ref{Zz1}) of entropy to temperature.
According to Eq. (\ref{3b}), definition of the compressibility
\begin{eqnarray}
\kappa\equiv\left({\partial x^{-1}\over\partial p}\right)^{-1}
\label{3c}
\end{eqnarray}
shows that it is reduced to the dynamic exponent (\ref{Q}).
Respectively, stability conditions
\begin{eqnarray}
C>0,\quad \kappa>0
\label{3d}
\end{eqnarray}
reduces to natural restriction
\begin{eqnarray}
z>1.
\label{3e}
\end{eqnarray}
It is principally important that the "pressure" (\ref{3a}) is introduced
as derivative of the specific entropy $h$
with respect to the microscopic scale $x$.
This means a microscopic nature of so defined "pressure", which
related to a susceptibility
(more exactly, the compressibility (\ref{3c})) of the time series with
respect to a choice of the microscopic scale $x$.

\section{Discussion}

We have addressed above the simplest model of the ideal gas, 
which have allowed us to examine analytically 
a self-similar time series in standard statistical manner. 
Characteristic peculiarity of related
equalities (\ref{I}) -- (\ref{K}), (\ref{O}) -- (\ref{S}),
(\ref{Zz}), (\ref{Zz1}), (\ref{S2}) -- (\ref{S4}),
(\ref{3b}), (\ref{3d}), (\ref{3e}) is a scale invariance with
respect to variation of the nonextensivity parameter $1-q$, which is
contained everywhere as combination (see Eq. (\ref{M})) 
\begin{eqnarray}
a=(1-q)C\quad C\equiv{DN\over 2}
\label{3x}
\end{eqnarray}
with free magnitude of parameter $a$ related to the
dynamic exponent $z$ (see Eq. (\ref{Q})). This invariance is clear
to be caused by self-similarity of the system under consideration.
For a given time series, the value of exponent $z$ is fixed if it
is addressed to a monofractal manifold and takes a closed set of
magnitudes $z$ in the case of self-similar system relevant to a
multifractal. In real time series this invariance can be broken,
so that a dependence on the parameter of nonextensivity could
appear. However, above introduced set of pseudo-thermodynamic
characteristics of time series is kept as applicable and
accustomed thermodynamic relations (\ref{1}), (\ref{1a}),
(\ref{3}), (\ref{1aa}), (\ref{3a}), (\ref{3c}) can be applied to
analysis of arbitrary time series.

Let us focus now on effect of external field and particle interaction, which switching on
is expressed by equalities (\ref{CC}), (\ref{CCC}). 
Related entropy additions (\ref{G2}), (\ref{G4}) arrive at 
total value of the specific heat (\ref{1aa}) in the following form:
\begin{eqnarray}
C_{tot}=C-{C\over T}\left(FX - {z-1\over Dz}{N\over V}w\right),
\label{Gg}
\end{eqnarray}
where one supposes $FV\ll T$ for simplicity. 
Thus, at temperature less than critical magnitude
\begin{eqnarray}
T_c=FxN^{1\over Dz} - {z-1\over Dz}w x^{-D}N^{1-{1\over z}}, 
\label{Gg1}
\end{eqnarray}
the system becomes nonstable due to external force $F$ and interparticle attraction $w<0$. 
If the number $N$ is much less than critical value $N_c$ defined as 
\begin{eqnarray}
N_c^{\left(1+{1\over D}\right){1\over z}-1}\equiv{z-1\over Dz}{|w|\over F}x^{-(D+1)}, 
\label{Gg2}
\end{eqnarray} 
main contribution gets interparticle attraction $w<0$, 
in opposite case $N\gg N_c$ -- external force $F$.
With accounting temperature definition (\ref{Ss}) and notation (\ref{P}), 
this instability means that a microscopic time interval $\tau$ should be more 
than a critical magnitude $\tau_c$.
In the case $N\ll N_c$, we find
\begin{eqnarray}
\tau_c={\left({z-1\over Dz}|w|\right)^{1\over 2(z-1)}\over\sqrt{2\pi{\rm e}T_0}}
x^{1-{D\over 2(z-1)}}N^{{1\over 2z}\left(1+{2\over D}\right)}.
\label{Gg3}
\end{eqnarray} 
In opposite case $N\gg N_c$, one obtains
\begin{eqnarray}
\tau_c={F^{1\over 2(z-1)}\over\sqrt{2\pi{\rm e}T_0}}
x^{2z-1\over 2(z-1)}N^{2z-1\over 2Dz(z-1)}.
\label{Gg3}
\end{eqnarray} 

On the other hand, the entropy additions (\ref{G2}), (\ref{G4}) 
accompanied by Eqs. (\ref{S1}), (\ref{3a}), (\ref{3b}) 
arrive at the compressibility (\ref{3c}) in the form
\begin{eqnarray}
\kappa^{-1}=z-(D+1){z-1\over z}{N\over X^D}\left(v+{w\over T}\right).
\label{333}
\end{eqnarray}
Hence, the system becomes nonstable if microscopic scale $x$ is less than a critical magnitude 
determined as
\begin{eqnarray}
x_c^D=(D+1){z-1\over z^2}\left(v+{w\over T}\right)N^{1-{1\over z}}.
\label{343}
\end{eqnarray}
In the limit of very large time intervals $\tau$, when the temperature is $T\gg w/v$, 
one finds the simple estimation
\begin{eqnarray}
x_c\sim\epsilon N^{{1\over D}\left(1-{1\over z}\right)}.
\label{344}
\end{eqnarray}
Thus, at $z=1$ critical value $x_c$ is reduced to a scale of the time series resolution 
$\epsilon$, being a particle radius within the gas model.

Above studied instabilities break the conditions (\ref{3d}), which guarantee  
predictability of the time series analysis. Our main result is that the predictability becomes  
inpossible if microscopic scales $\tau$, $x$ take values less than critical magnitudes $\tau_c$, $x_c$, 
increasing with growth of the number $N$ of time series terms.

To conclude, we point out that main advantage of the approach proposed 
is a possibility of its application
to numerical analysis of real time series. The basis of this calculations
is stated on expressions (\ref{A}) -- (\ref{Ccc}), (\ref{D}) -- (\ref{F}), (\ref{Ss}),
(\ref{1}), (\ref{1a}), (\ref{3aa}), (\ref{3}), (\ref{1aa}), (\ref{S1}), (\ref{3a}), (\ref{3c}),
(\ref{H1}). This work is in progress.

\section*{Acknowledgments}\label{sec:level1}

Author is grateful to Dr. H. Kantz for nice introduction into the time series
analysis and helpful discussions. 
I am obliged also to Dr. D. Kharchenko for assistance.

\section*{Appendix: Non-extensive thermodynamics of thermodynamic system}\label{sec:level2}

Calculation of main thermodynamic quantities of nonextensive
ideal gas arrives at the Euler $\Gamma$-function to be reduced
to the following expressions for partition function and internal
energy \cite{9} -- \cite{7}
\begin{eqnarray}
Z={V^N\gamma(q)\over N!}\left[{\theta\left<1\right>_q\over 1-q}\right]^{DN\over
2}\left[1+(1-q){DN\over 2}\right]^{-1}
\left[1+(1-q){~E\over\left<1\right>_q T_0}\right]^{{1\over 1-q}+{DN\over 2}},
\label{S5}
\end{eqnarray}
\begin{eqnarray}
E={DN\over 2}{V^N\gamma(q)T_0\over N!}\left[{\theta\left<1\right>_q\over
1-q}\right]^{DN\over 2}
\left[1+(1-q){DN\over 2}\right]^{-1}
\left[1+(1-q){~E\over\left<1\right>_q T_0}\right]^{{1\over 1-q}+{DN\over
2}}Z^{-q}.
\label{S6}
\end{eqnarray}
Here, $D$-dimensional gas in volume $V\equiv X^D$ is
addressed and the notations are introduced 
\begin{eqnarray}
\theta\equiv{2\pi m T_0\over h^2},\quad
\gamma(q)\equiv{\Gamma\left({1\over 1-q}\right)\over\Gamma\left({1\over
1-q}+{DN\over 2}\right)},
\label{S7}
\end{eqnarray}
where $m$ is particle mass, $h$ is Planck constant and $T_0$ is energy
scale.
Normalization quantity
\begin{eqnarray}
\left<1\right>_q=Z^{1-q}
\label{G}
\end{eqnarray}
follows from the normalization condition of the Tsallis' distribution
function \cite{2}.

Switching on an external force ${\bf F}={\bf const}$ causes the second term 
in Hamiltonian (\ref{CC}) to arrive at the following factor in Eq. (\ref{S5}):
\begin{eqnarray}
Z_{ext}=\exp\left[DN\left({FX\over 2T}\right)\right]
\left[{\sinh\left({FX\over 2T}\right)\over\left({FX\over 2T}\right)}\right]^{DN}.
\label{G1}
\end{eqnarray}
According to the definition (\ref{H1}), this yields the entropy addition
\begin{eqnarray}
H_{ext}= DN\left({FX\over 2T}\right)
+DN\ln\left[{\sinh\left({FX\over 2T}\right)\over\left({FX\over 2T}\right)}\right].
\label{G2}
\end{eqnarray}
Within a supposition $F>0$, a homogeneous external field arrives at monotonous decrease 
of the entropy $H_{ext}(T)$ with the temperature growth. 
At $T\gg FX$ contribution of the second term is neglecting, 
in the limit $T\to 0$ it tends to the first one.

According to recent work \cite{11}, cluster expansion of 
the particle interaction in Hamiltonian (\ref{CCC}) 
results in additional factor in partition function (\ref{S5}): 
\begin{eqnarray}
Z_{int}= 1 - {N^2\over 2V}\left(v+{w\over T}\right);\quad
w\equiv S_D\int\limits_\epsilon^\infty w(x)x^{D-1}{\rm d}x,~
S_D\equiv{2\pi^{D/2}\over\Gamma(D/2)};~~
v\equiv{S_D\over D}\epsilon^D,
\label{G3}
\end{eqnarray}
where $\epsilon$ is effective radius of the particle core. 
Relevant entropy addition
\begin{eqnarray}
H_{int}=-{aN^2\over 2V}\left(v+{w\over T}\right)
\label{G4}
\end{eqnarray}
monotonously decreases with growth of the particle volume $v$. Growth of the temperature $T$ 
causes the entropy decrease for attractive interaction $w<0$ and its increase 
in the case of repelling one $(w>0)$.

Finally, we arrive at useful relations between physically defined entropy $H$ and 
related Tsallis' magnitude $H_q$. One has symbolic equation
\begin{eqnarray}
H=a\ln[\exp_q(H_q)],\quad H_q=\ln_q[\exp(H/a)];\qquad a\equiv (1-q){DN\over 2},
\label{Aa}
\end{eqnarray}
that are alternations of the usual functions logarithm $\ln(x)$ and
exponential $\exp(x)$ with corresponding Tsallis' generalizations
\cite{2}
\begin{eqnarray}
\ln_q(x)\equiv{x^{1-q} - 1\over 1-q},\quad
\exp_q(x)\equiv[1+(1-q)x]^{1\over 1-q}.
\label{00}
\end{eqnarray}
In explicit form, relations (\ref{Aa}) are appeared as
\begin{eqnarray}
H\equiv a\ln Z,\quad
H_q\equiv{\left< 1\right>_q - 1\over 1-q}.
\label{H1}
\end{eqnarray}

\end{document}